\documentclass[twocolumn]{aastex61}
\hypersetup{linkcolor=red,citecolor=blue,filecolor=cyan,urlcolor=magenta}

\received{2017 Decembber 17}
\revised{2018 March 3}
\accepted{2018 March 6}
\submitjournal{ApJ Letter}

\shorttitle{ High-frequency Oscillations in the Atmosphere above a Sunspot Umbra }
\shortauthors{Wang et al.}

\begin{document}

\title{High-frequency oscillations in the atmosphere above a sunspot umbra}

\correspondingauthor{Song FENG}
\email{feng.song@astrolab.cn }

\author{Feng WANG}
\affiliation{School of Physics and Electronic Engineering, Guangzhou University, Guangzhou 510006, China}
\affiliation{Yunnan Key Laboratory of Computer Technology Application/Faculty of Information Engineering and Automation, Kunming University of Science and Technology, Kunming 650500, China;}

\author{Hui DENG}
\affiliation{School of Physics and Electronic Engineering, Guangzhou University, Guangzhou 510006, China}

\author{Bo LI}
\affiliation{Shandong Provincial Key Laboratory of Optical Astronomy and Solar-Terrestrial Environment, Institute of Space Sciences, Shandong University, Weihai 264209 China;}

\author{Song FENG$^\dag$}
\affiliation{School of Physics and Electronic Engineering, Guangzhou University, Guangzhou 510006, China}
\affiliation{Yunnan Key Laboratory of Computer Technology Application/Faculty of Information Engineering and Automation, Kunming University of Science and Technology, Kunming 650500, China;}

\author{Xianyong BAI}
\affiliation{CAS Key Laboratory of Solar Activity, National Astronomical Observatories, Beijing 100012, China}

\author{Linhua DENG}
\affiliation{Yunnan Observatories, Chinese Academy of Sciences, Kunming 650216, China}

\author{Yunfei YANG}
\affiliation{Yunnan Key Laboratory of Computer Technology Application/Faculty of Information Engineering and Automation, Kunming University of Science and Technology, Kunming 650500, China;}

\author{Zhike XUE}
\affiliation{Yunnan Observatories, Chinese Academy of Sciences, Kunming 650216, China}

\author{Rui WANG}
\affiliation{Yunnan Observatories, Chinese Academy of Sciences, Kunming 650216, China}

%
\begin{abstract}

We use high spatial and temporal resolution observations, simultaneously obtained with the New Vacuum Solar Telescope and Atmospheric Imaging Assembly (AIA) on board the \textit{Solar Dynamics Observatory}, to investigate the high-frequency oscillations above a sunspot umbra. A novel time--frequency analysis method, namely the synchrosqueezing transform (SST), is employed to represent their power spectra and to reconstruct the high-frequency signals at different solar atmospheric layers. A validation study with synthetic signals demonstrates that SST is capable to resolving weak signals even when their strength is comparable with the high-frequency noise. The power spectra, obtained from both SST and the Fourier transform, of the entire umbral region indicate that there are significant enhancements between 10 and 14 mHz (labeled as 12 mHz) at different atmospheric layers. Analyzing the spectrum of a photospheric region far away from the umbra demonstrates that this 12~mHz component exists only inside the umbra. The animation based on the reconstructed 12 mHz component in AIA 171 \AA\ illustrates that an intermittently propagating wave first emerges near the footpoints of coronal fan structures, and then propagates outward along the structures. A time--distance diagram, coupled with a subsonic wave speed ($\sim$ 49 km s$^{-1}$), highlights the fact that these coronal perturbations are best described as upwardly propagating magnetoacoustic slow waves. Thus, we first reveal the high-frequency oscillations with a period around one minute in imaging observations at different height above an umbra, and these oscillations seem to be related to the umbral perturbations in the photosphere. 

\end{abstract}



\section{Introduction} 

Periodic oscillations abound in the atmosphere of sunspots as revealed by the temporal variations in, say, Doppler velocities and intensities. These oscillations are usually considered as slow magnetoacoustic waves. The magnetic field orientation is different in the umbra from that in the penumbra; however, the propagation of slow magnetoacoustic waves is anisotropic to the orientation of magnetic fields. So, different types of magnetohydrodynamic waves are displayed inside and above sunspots at different solar atmospheric layers, such as oscillations above light bridges \citep{2013ApJ...767..169L,2013A&A...560A..84S,2014ApJ...792...41Y, 2015MNRAS.452L..16B, 2016ApJ...816...30S,  2016A&A...594A.101Y}, umbral flashes \citep{2014RAA....14.1001F}, penumbral running waves \citep{2000A&A...354..305C,2013ApJ...779..168J,2015A&A...580A..53L} and coronal fan structures \citep{2011A&A...533A.116Y,2012ApJ...757..160J,2016ApJ...823L..16T}. Slow magnetoacoustic waves are assumed to result from the interaction between the photospheric $p$-modes and magnetic fields, and can propagate upward along magnetic field lines, and finally reach coronal heights \citep{2012ApJ...757..160J,2016NatPh..12..179J, 2015ApJ...812L..15K, 2017ApJ...847....5K,2008ApJ...682L..65Y}. Studies on the multiple oscillations above the sunspot atmosphere prove useful for understanding the physical conditions of oscillations and waves, and their formation and propagation mechanisms \citep{2015LRSP...12....6K,2016ApJ...830L..17Z,2016NatPh..12..179J}. 

The frequency distributions of sunspot oscillations usually tend to be concentrated in the bandwidth of 3--6 mHz (i.e., 3--5 minutes). It is believed that three-minute umbral oscillations exist in the chromosphere, and five-minute oscillations exist in the photosphere due to a cutoff frequency. Recently, thanks to the high spatial and temporal observations acquired with ground- and space-based instruments, it has been found that the oscillations below the typical acoustic cutoff frequency can also reach chromospheric heights, and even propagate into the corona along the magnetic field lines. For example, \citet{2014ApJ...792...41Y} detected five-minute oscillations at a light bridge observed at chromospheric heights. Employing a multi-wavelength approach, \citet{2012ApJ...757..160J} investigated three-minute magnetoacoustic waves, and revealed the propagation from the photosphere, through the chromosphere, and into the corona. \citet{2012ApJ...746..119R} studied the perturbations in the 5--9 mHz frequency range in different umbra atmospheres, and examined its propagation along inclined magnetic field lines. \citet{2013A&A...554A.146K} found a high-frequency mode with 8 mHz (2 minutes) located at a peculiar location inside an umbra, and concluded that the frequency probably relates to the oscillation of umbral dots in the photosphere. \citet{2014A&A...569A..72S} further revealed that the spectra of umbral oscillations contain distinct peaks at 1.9, 2.3, and 2.8~minutes, and the oscillations in the higher atmospheric layers occur later than in lower ones, and claimed that they are upward propagating waves. 

So far, the high-frequency perturbation modes around one minute have been reported at a particular atmosphere height \citep{2016A&A...594A.101Y,2017ApJ...847....5K}. Utilizing the multiple optical and UV wavelength observations obtained by the Dunn Solar Telescope (DST) and the \textit{Interface Region Imaging Spectrograph (IRIS)}, \citet{2016A&A...594A.101Y} studied the oscillations in the emission intensity of the light bridge plasmas at different temperatures, and revealed that the light bridge exhibits a perturbation with some period shorter than one minute in a particular channel at the solar atmosphere. 
\citet{2017ApJ...847....5K} confirmed that a high-frequency oscillation mode with 13.1~mHz is located near an umbral center in the DST Ca \textsc{ii} K and \textit{IRIS} 2796 \AA\ channels, and does not seem to be related to the magnetic field inclination angle effects. A similar period is also found in a flare loop observed by the \textit{Geostationary Operational Environmental Satellites} \citep{Ning:2017eh}. However, a periodicity around one minute has not been found in imaging observations at other umbral atmospheric layers. 

In this letter, we use ground- and space-based imaging instruments, with high spatial and temporal resolution, to investigate the high-frequency mode with a period around one minute at different heights above a sunspot umbra. For representing, decomposing, and reconstructing high-frequency oscillation modes, we employ a novel time--frequency analysis technique named synchrosqueezing transform \citep[SST;][]{Daubechies:2011bg,Thakur:2013kc}. This technique has been used \citep{2013ApJ...771...33S,2017ApJ...845...11F} to represent intrinsic modes of solar cycles. Furthermore, \citet{2017ApJ...845...11F} demonstrated that it can decompose and reconstruct intrinsic modes with a high spectral resolution without mode mixing. In Section~\ref{sec_valid}, we first conduct a simulation experiment to evaluate the performance of SST. We then describe, in Section~\ref{sec_data}, our observations and data processing procedures as well as an error analysis. Finally, Section~\ref{sec_disc} concludes this study.  

\begin{figure*}
	\centering
        \includegraphics[width=18cm]{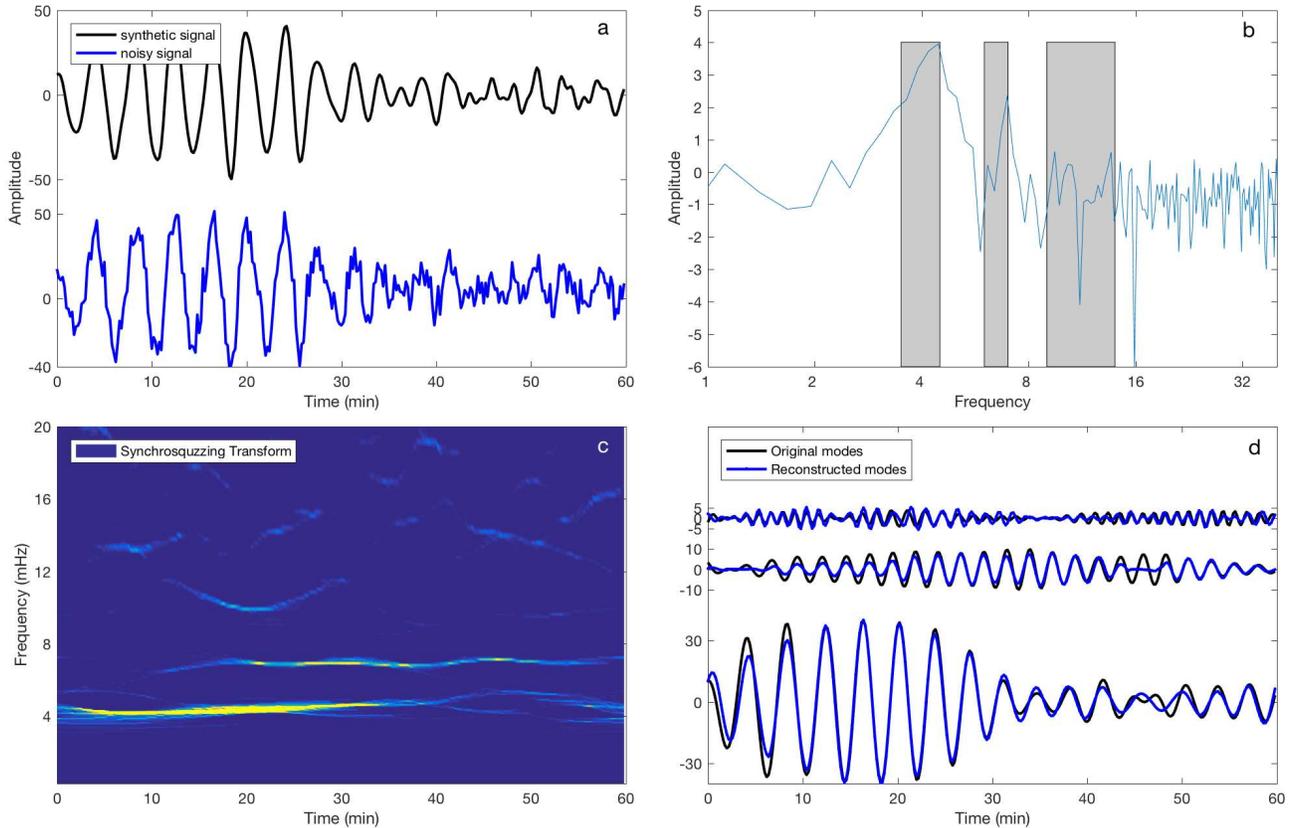}
	\caption{Simulation experiment for evaluating the performance of SST. (a) The synthetic signal without noise (black curve) and with noise (blue curve). The noise follows a Poisson distribution. Its standard error is 6.3, and the S/N ratio is about 3. The synthetic signal consists of three frequency components within 3--4, 6--7, and 9--14 mHz, and their amplitudes are approximately 40, 10, and 4. These are generated by a random process. (b) Power spectrum of Fourier transform. Gray regions denote the frequency ranges of the three components. (c) Power spectrum of the SST. (d) Original components (black curves) and the ones reconstructed from the noisy signal (blue curves).}
		\label{fig1}
\end{figure*}

\section{Performance evaluation of the synchrosqueezing transform}
\label{sec_valid}

SST is a time--frequency analysis technique based on the continuous wavelet (CWT) and the spectral reassignment method. The reassignment method compensates for the spreading effects inherent to CWT. SST concentrates the frequency content only along the frequency direction, and preserves the time resolution of a quasi-periodic signal. Moreover, the inverse synchrosqueezing algorithm can reconstruct instantaneous frequencies of the signal. More detailed descriptions on SST can be found in \citet{Daubechies:2011bg} and \citet{Thakur:2013kc}. 

For evaluating the performance of SST, we synthesize a signal $s(t)$ with three frequency components, and each component varies around the central values of 4, 6, and 12 mHz along the time direction, respectively. Moreover, their amplitudes also randomly fluctuate around specific values. The amplitudes of the 4, 6, and 12 mHz components are approximately 40, 10, and 4, respectively. It should be noted that all the frequency and amplitude fluctuations are generated by a random process around the given value for a near ``real'' signal. 

For an observational image, its counting noise is mainly generated by Poisson processes. So, a Poisson noise with a signal-to-noise ratio (S/N) being 3 and standard error being 6.3 is considered. Figure 1a shows the signal we actually analyze (blue) and its noise-free counterpart (black). In the noise-free signal, the power is located in the ranges with 3--4 mHz, 6--7 mHz, and 9--14 mHz, respectively. The sampling interval and the total number of data points are 12 s and 300, respectively. The amplitude of the 12 mHz component is only 10 \% of the 4 mHz component. In absolute values, this amplitude ($\sim 4$) is less than the standard error ($\sim 6.3$) of the Poisson noise. 

\begin{figure*}
	\centering
        \includegraphics[width=8cm]{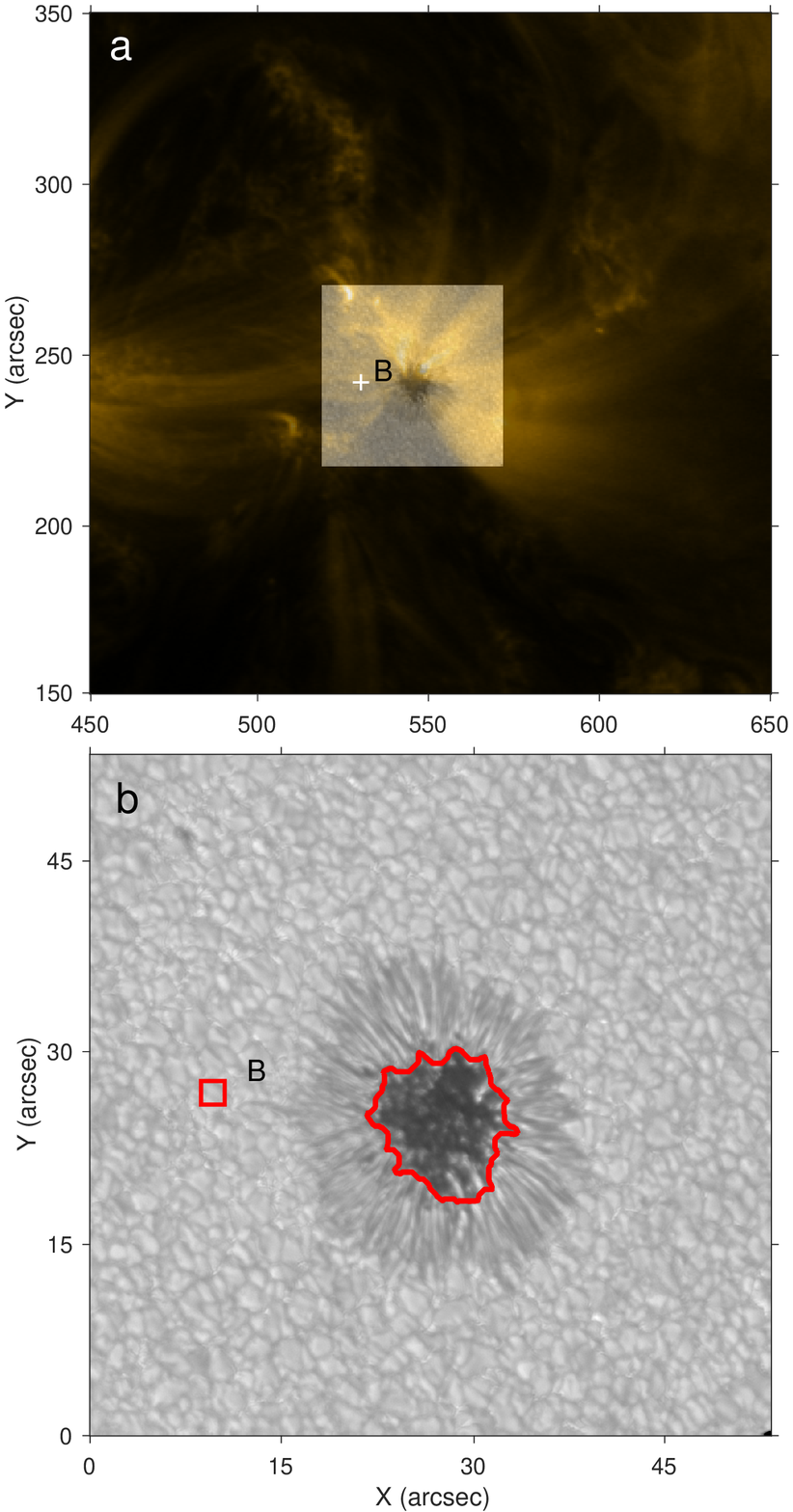}
        \includegraphics[width=8cm]{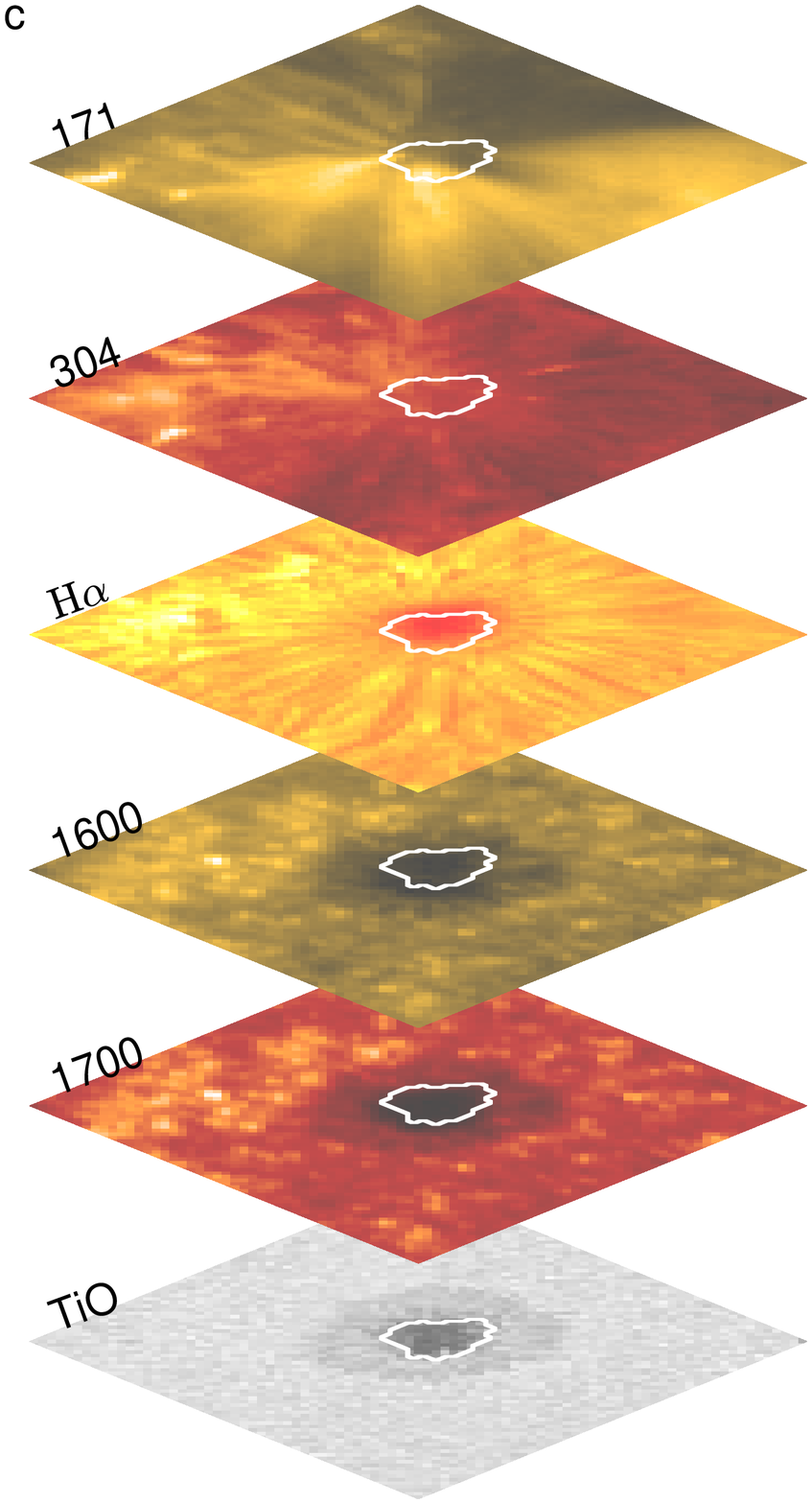}
	\caption{(a) An AIA 171 \AA\ image. (b) The co-temporal and co-spatial NVST TiO image, which is also superimposed on the 171\AA\ image. The pixel size of TiO images is 0\arcsec.041. The umbral region is outlined by a red curve. A region marked ``B" is used to compare the spectra to the umbral region in Section 4. (c) A co-spatial stacked image. From bottom to top, the images represent observations from a variety of ground- and space-based instruments: NVST TiO, AIA 1700 \AA, AIA 1600 \AA, NVST H$\alpha$, AIA 304 \AA, and AIA 171 \AA. The white curves mark the umbral region that is the region of interest in this work.}
	\label{fig2}
\end{figure*}

To evaluate the spectrum representation performance based on SST, the power spectra of the noisy signal (the blue curve in Figure 1a) based on SST and the Fourier transform are shown in Figures 1c and 1b, respectively. The gray regions in Figure 1b denote the ranges of the three frequency components. One can see that the noise floor appears at about 8 mHz, and the component between 9 and 14 mHz is heavily contaminated by the high-frequency noise. In Figure 1c, there are some spectral contents, although weak, in the range 9--15 mHz, demonstrating that the SST has better ability to suppress random noise.  Figure 1d shows the original frequency components (black) and the corresponding reconstructed modes (blue). Compared with the original components, the amplitudes of the reconstructed modes are slightly different. However, both the frequency and the phase are almost the same as in the original ones, which is particularly true for the weakest component (12 mHz). To sum up at this point, SST performs well for resolving the physical signals that are heavily contaminated by a Poison noise.

\begin{figure*}
	\centering
        \includegraphics[width=18cm]{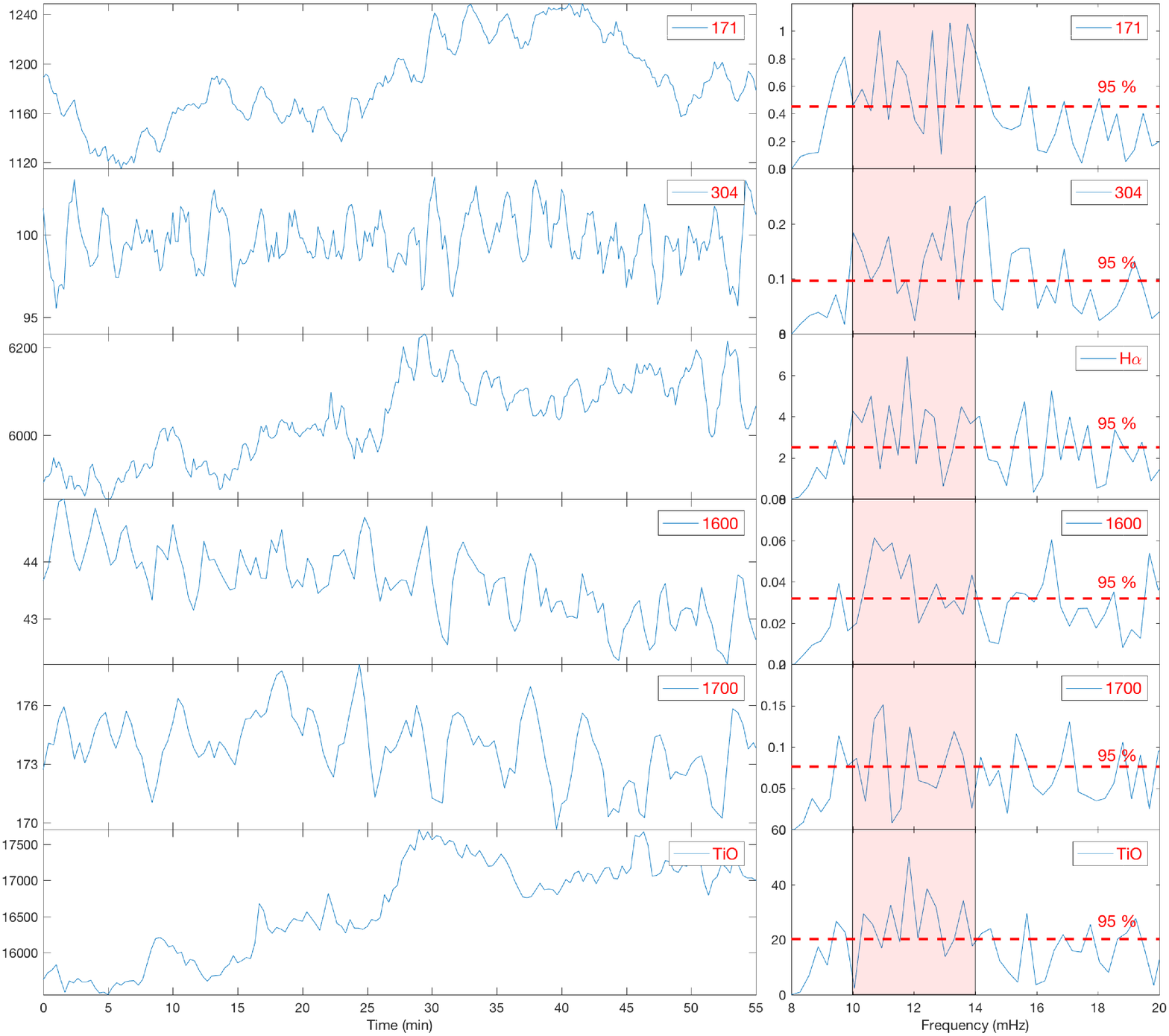}
\caption{Left column: intensity curves averaged over the entire umbra region in different channels. Right column: the corresponding Fourier spectra of the de-trended curves. Red dashed curves denote the 95 \% confidence levels. The shaded regions denotes the frequency range of interest in this work.}
		\label{fig3}
\end{figure*}

\section{Observations and Data Reduction}
\label{sec_data}

\begin{figure*}
	\centering
        \includegraphics[width=18cm]{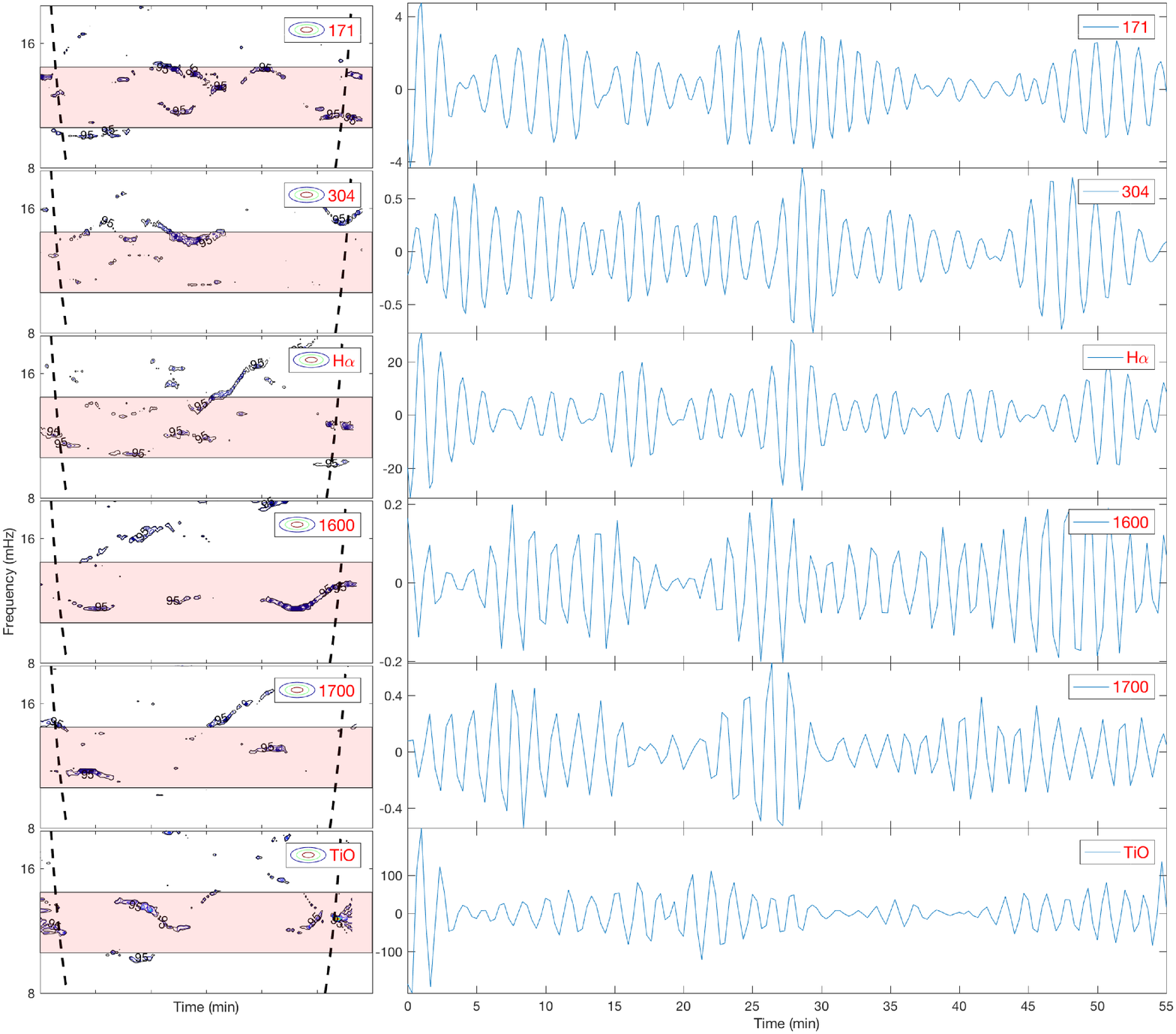}
\caption{Left column: SST power spectra in each channel. Black dashed curves denote the influence of cone, and the contours present 95 \% confidence levels in the 12 mHz components. Right column: the corresponding reconstructed components. }
		\label{fig4}
\end{figure*}

This study focuses on active region NOAA 11081, located in the northern hemisphere on 2013 August 1. The New Vacuum Solar Telescope \citep[NVST;][]{2014RAA....14..705L,2014IAUS..300..117X}, located at the Fuxian Solar Observatory of the Yunnan Observatories, China, was employed to obtain the high-resolution ground-based imaging observations. The two optical channels, centered at the H$\alpha$ line core (6562.8$\pm$0.25 \AA) and TiO (7058$\pm$10~\AA), were used to capture the chromospheric and photospheric images between 03:38 and 04:35 UT. The pixel scale is 0\arcsec.162 for the H$\alpha$ images, and 0\arcsec.041 for the TiO ones. The sampling interval is 12 s for the H$\alpha$ sequences, and 20 s for the TiO ones. 

\begin{figure}
	\centering
        \includegraphics[width=8cm]{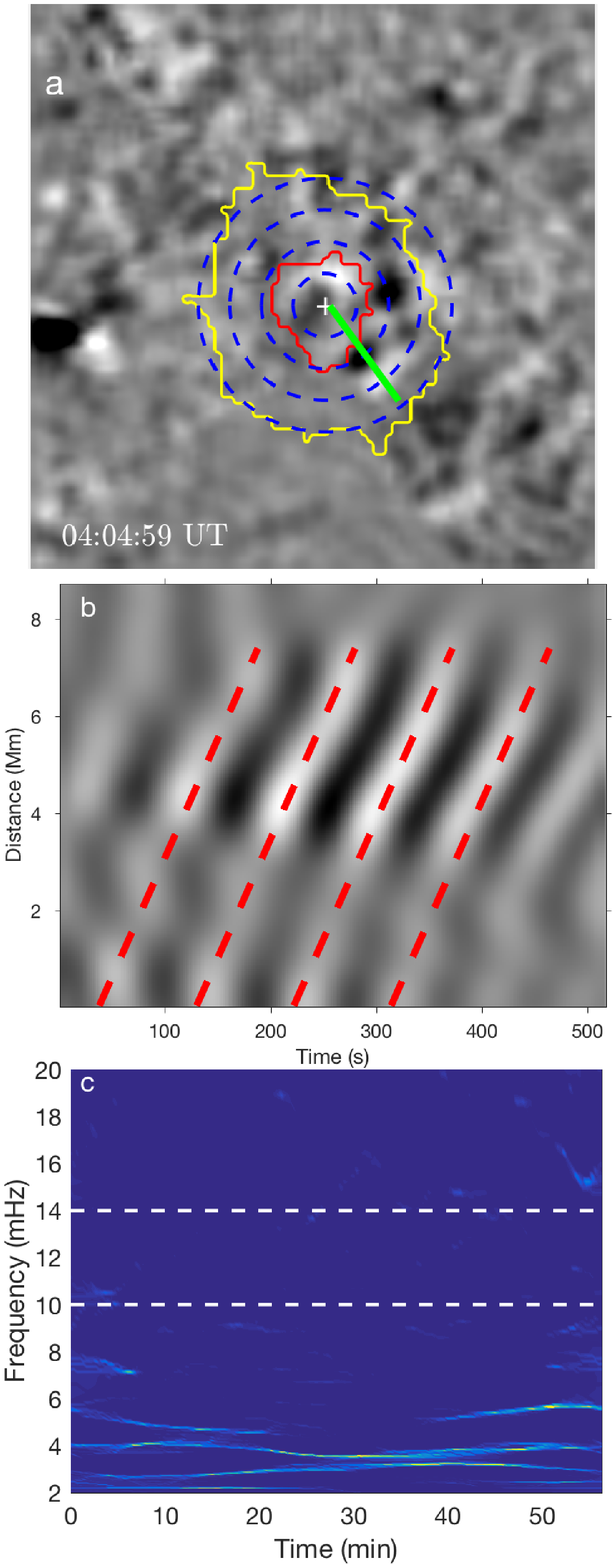}
	\caption{(a) A reconstructed running difference based on every pixel in the AIA 171 \AA\, whose spectra only includes the frequency contents from 10 to 14 mHz. Its original image is observed on 2013 August 1 at 04:04:59 UT. The red and yellow curves outline the umbra and penumbra edge, and the dashed blue lines highlight expanding annulus. A cut taken to make the time--distance diagram is indicated with a green line. (b) The time--distance diagram started at 04:00:35 UT and finished at 04:08:59 UT covers about six cycles of the propagating features. Its propagating velocity is about 49 $\pm$ 4 km s$^{-1}$. (c) The power spectrum of region ``B" marked in Figures 2a and b. Two horizontal dashed lines indicate the range between 10 and 14 mHz. }
		\label{fig5}
\end{figure}

The co-temporal and co-spatial observations were also obtained from the Atmospheric Imaging Assembly \citep[AIA;][]{2012SoPh..275...17L} on board the \textit{Solar Dynamics Observatory} \citep[SDO;][]{2012SoPh..275....3P}. Our analysis concentrates on the 171 \AA, 304 \AA, 1600 \AA\, and 1700 \AA\ channels. The formation height of the 1700 \AA\ channel is located at the temperature minimum, 1600 \AA\ at the lower chromosphere, 304 \AA\ at the transition region, and 171 \AA\ at the corona. To get the level 1.5 data, the level 1.0 data were processed using the routine $aia\_prep.pro$. The pre-process brings all the data to a common center and plate scale with $\approx$ 0\arcsec.6, and the cadence is 12 s for the AIA 171\AA\ and 304 \AA\ channels, and 24 s for the 1600 \AA\ and 1700 \AA. Furthermore, the images were co-aligned by the sub-pixel registration algorithm \citep{Feng:2012hk}. Similarly, all the NVST images were aligned to the AIA data with a scale reduction and a rotation transform. Finally, we truncated the size of the data sets into a set of 54\arcsec\ $\times$ 54\arcsec, which contains the entire sunspot. 

Figure 2a shows an AIA 171 \AA\ image and its co-temporal and co-spatial TiO thumbnail. The original TiO image is shown in Figure 2b. The umbral edge is marked with a red curve. The entire umbra is employed to analyze high-frequency oscillation modes. The images stacked in Figure 2c represent the observations at different formation heights.

Figure~3 presents the average intensity variations of the entire umbra in every channel (left column).  We removed long-term trend greater than 2.5 minutes in all data with frequency filtering for concentrating on high-frequency component analysis,  and the power spectra of the Fourier transform of the de-trended curves are presented in the right column. Red dashed lines indicate the 95\% confidence levels, and the shaded regions denote a frequency range between 10 and 14 mHz. We can find that the spectral contents between 10 and 14 mHz exist in each  channel, and are higher than the 95 \% confidence level. There are significant peaks within the shaded regions.  In what follows, we use ``the 12~mHz component'' to label the oscillations in this frequency range for brevity. Subsequently, the 12~mHz components are represented and reconstructed by the SST, respectively.  The SST spectra are represented in the left column of Figure 4, in which the black dashed curves denote the cone of influence and those contours present 95 \% of the confidence level. The shaded regions are decomposed and reconstructed by the SST and are shown in the right column of Figure 4.

For estimating the amplitude error of the reconstructed components, we analyze their noise level. The error estimation proposed by \citet[see Section 3 by][]{2012A&A...543A...9Y} is used to calculate the image noise in each AIA channel. For a pixel with an intensity value $F$ (in units of DN), the standard error of the noise can be approximated by $\sqrt{2.3+0.06F}$. Because the light curve in the AIA 171 \AA\ channel incorporates the entire umbra region that amounts to 190 pixels in total, the estimated data noise should be multiplied by a factor of 1/$\sqrt{190}$. Thus, the error of the AIA 171 \AA\ channel is about 0.6~DN for the selected data. The error can be estimated similarly for the rest of the EUV/UV channels. 

For the NVST data, we assume that the noise is primarily due to photon counting, which follows a Poisson distribution. Therefore, its amplitude is the square root of the signal intensity. Because the pixel resolution of the original image of the TiO data is reduced from 0\arcsec .041 to 0\arcsec .6, and a 190-pixel region is selected to generate the light curve, the noise is about 1/202 (1/50 for the H$\alpha$ channel) of the square root of the original intensity. The end result is that the amplitude error of the TiO channel is about 0.6, and that of the H$\alpha$ is about 1.5. So, for AIA 1700, 1600, and 304 \AA, the amplitudes of the reconstructed mode are two times larger than the estimated noise. Likewise, for AIA 171 \AA, the NVST TiO, and H$\alpha$, the amplitudes are at least three times larger. From this error analysis, together with the confidence levels in the Fourier and SST spectra, we conclude that the reconstructed components of 12 mHz are physical rather than high-frequency noise. 

\begin{figure*}
	\centering
        \includegraphics[width=0.9\textwidth]{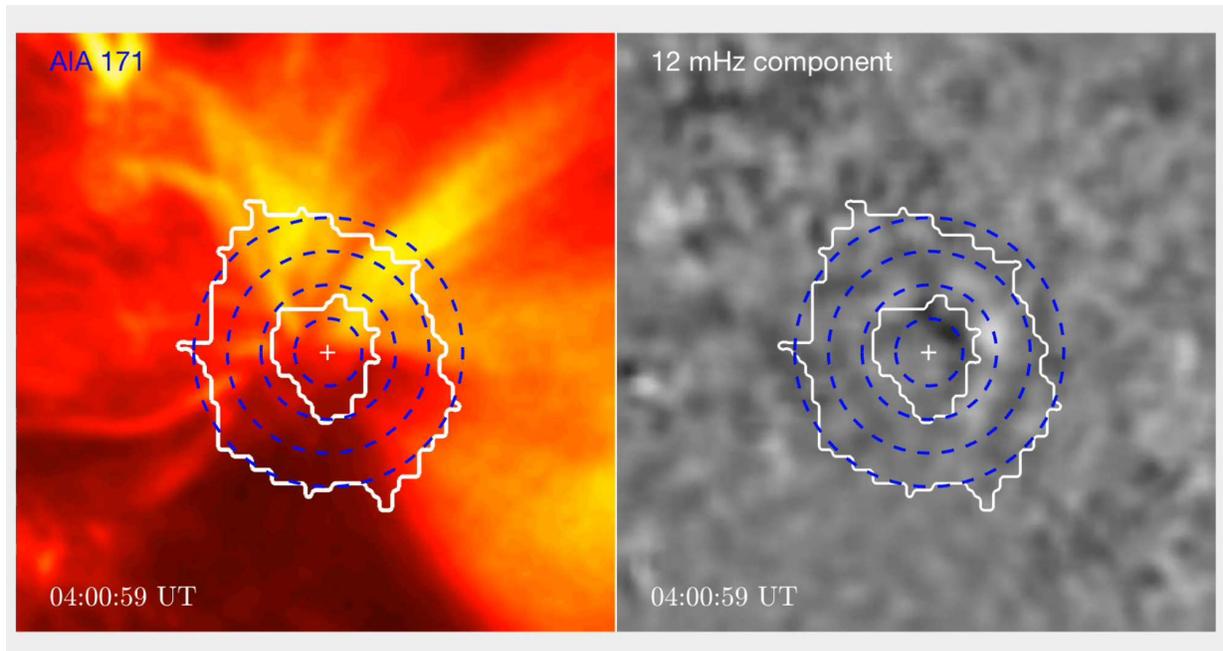}
	\caption{Online animation of the the AIA 171 \AA\ images (left) and the 12~mHz modes reconstructed by every pixel in the AIA 171 \AA\ channel (right).}
		\label{fig6anim}
\end{figure*}

We reconstructed the 12~mHz modes for every pixel in the AIA 171 \AA\ channel, and then generated a running-difference animation (Figure 6). This animation is restricted to the time interval between 03:59:59 and 04:09:47 UT, during which the 12~mHz signal is the most evident (see the uppermost panel in Figure~4). In the animation, the footpoint of the coronal fan structures is marked with a white cross. The umbral and penumbral borders as determined with the NVST TiO images are marked with the red and yellow curves, respectively. The dashed blue lines highlight a series of expanding annuli. The movie exhibits an intermittent and outward propagating wave near the footpoint of the coronal fan structures. It emerges first near the footpoint and then propagates outward along the fan structures before disappearing. A running-difference image, extracted from the animation, is shown in Figure 5a, where we also display an artificial slit extending from the footpoint and following the coronal fan structure (the thick green line). A time--distance plot along this slit, 24 pixels long and 3 pixels wide, is shown in Figure 5b. The propagating perturbations are featured by the repeating diagonal ridges, which we outline with a series of red dashed lines to help guide the eye. Standard practice yields that the disturbances propagate at a speed of 49 $\pm$ 4 km s$^{-1}$.

To further understand where the 12~mHz signal in the TiO image sequences comes from, we also examined the power spectrum in the TiO observations for the region in the quiet photosphere that surrounds the sunspot. Shown in Figure 5c is a spectrum at an arbitrarily chosen location (marked ``B'' in the left column of Figure~2). Evidently, this spectrum does not exhibit any power in the range of 10--14 mHz, delineated by the horizontal dashed lines. This means that the 12~mHz signal originates from within the umbra rather than coming from the surrounding quiet photosphere. 

\section{discussion and conclusion}
\label{sec_disc}

A high-frequency oscillation with a frequency range 10--14 mHz (about one minute) above a sunspot umbra atmosphere was found using simultaneous observations captured by the NVST and SDO/AIA instruments. A novel time--frequency analysis method, called SST, was used to decompose and reconstruct the high-frequency components. To ensure that the reconstructed components are physical rather than high-frequency noise, a synthetic signal (the blue curve in Figure 1a), the noise floor appears at even lower frequencies (the spectral break now occurs at 8 mHz) was analyzed. The results decomposed and reconstructed demonstrate that the SST has no problem in resolving the physical signal even if the contamination from high-frequency noise is even stronger in the synthetic time series. Analyzing Fourier and SST spectra of the intensity curves at six channels: the AIA 171, 304, 1600, 1700 \AA\, the NVST H$\alpha$, and TiO, the spectral contents between 10 and 14 mHz not only have significant peaks, but also higher than 95 \% confidence levels.  Furthermore, error analysis demonstrates that all the reconstructed amplitudes of the 12 mHz signals are at least twice larger than the corresponding error. The time--distance diagram, coupled with a subsonic speed ($\sim$ 49 km s$^{-1}$), highlights the fact that these coronal perturbations are best described as upwardly propagating magnetoacoustic slow-mode waves.

Compared with the spectra from the region surrounding the sunspot, the 12~mHz periodicity exists only inside the umbra, meaning that the modes at the coronal height are probably related to the perturbations from the photospheric umbra. We note that \citet{2017ApJ...847....5K} have already found a similar periodicity at chromospheric heights above an umbra. However, to our knowledge, the present study is the first to reveal the existence of such a periodicity at different heights above an umbra in imaging observations.  

Before closing, we would like to point out that our findings benefit from the capability of SST for suppressing spectral smearing. As such, SST is likely to find some wider use for representing and decomposing intrinsic modes in non-stationary signals.

\acknowledgments
The authors thank the referee for useful comments. We thank the NVST team for their high-resolution observations and the use of \textit{SDO/AIA} image obtained courtesy of NASA/SDO and the AIA science teams. This work is supported by the National Natural Science Foundation of China (Numbers: 11463003, 11503080, 11761141002, 41474149, 41674172) and the Joint Research Fund in Astronomy (U1531132, U1531140, U1631129) under coperative agreement between the NSFC, the National Key Research and Development Program of China (Number: 2016YFE0100300), and the CAS Key Laboratory of Solar Activity of National Astronomical Observatories (KLSA201715).

%


\end{document}